\def\erg{{\rm\thinspace erg}}

\def\K{{\rm\thinspace K}}

\def\km{{\rm\thinspace km}}
\def\kpc{{\rm\thinspace kpc}}

\def\Mpc{{\rm\thinspace Mpc}}

\def\s{{\rm\thinspace s}}
\def\yr{{\rm\thinspace yr}}

\def\ergps{\hbox{$\erg\s^{-1}\,$}}

\def\kmps{\hbox{$\km\s^{-1}\,$}}

\def\kmpspMpc{\hbox{$\kmps\Mpc^{-1}$}}
\def\etal{\hbox{et al.}}

\documentclass{ws-procs9x6}

\begin{document}

\title{X-rays from Clusters of Galaxies}

\author{A.C. Fabian and S.W. Allen}

\address{Institute of Astronomy, \\
Madingley Road, \\
Cambridge CB3 0HA, \\
U.K.\\
E-mail: acf,swa@ast.cam.ac.uk}


\maketitle

\abstracts{The X-ray emission from clusters of galaxies enables them to
be used as good cosmological probes and as an example for massive
galaxy formation. The gas mass fraction in clusters should be a
universal standard which by means of Chandra observations enables
$\Omega_{\rm m}$ to be determined to better than 15 per cent accuracy.
Future observations of its apparent variation with redshift will
enable $\Omega_{\rm \Lambda}$ to be measured. The interplay of
radiative cooling and heating in cluster cores may reveal the dominant
processes acting during the formation of the baryonic part of massive
galaxies. }

\section{Introduction}

Clusters of galaxies are luminous X-ray sources, with X-ray luminosities
ranging from $10^{43}- 10^{46}\ergps$. The emission is
predominantly thermal bremsstrahlung from hydrogen and helium in the
intracluster medium. Line emission, particularly from iron, is also
present showing that most of the gas has a mean metallicity of about 0.3
Solar. The total mass of the intracluster medium is about one tenth 
of the total cluster mass, and about 6 times that of all the stars in
the member galaxies. Most of the mass of a cluster is due to dark
matter.

Clusters are the most massive bound objects in the Universe and
therefore make good cosmological probes. They are the extreme tail of
the mass distribution. The number density of clusters in a given mass
range is a sensitive measure of the amplitude of the cosmic power
spectrum on cluster scales, $\sigma_8$. The gas fraction in clusters,
$f_{\rm gas}$, enables the matter density parameter, $\Omega_{\rm M}$,
to be determined and is a standard measure which should be invariant
with redshift. This means that it has a strong potential to be a
valuable, independent diagnostic of dark energy, $\Omega_{\rm \Lambda}$.

The X-ray emission in the cores of many clusters is sharply peaked.
The radiative cooling time of the gas within 50~kpc of the centre is
shorter than the likely age of the cluster. The temperature drops
smoothly there by a factor of two to three from that of the outer gas.
Although it might seem that a cooling flow should be operating there
with gas cooling out of the intracluster medium, spectra from the new
generation of X-ray space observatories show that radiative cooling is
much reduced and some form of distributed heating is taking place.

This last point is of considerable importance for understanding the
gaseous part of galaxy formation, most of which proceeds by radiative
cooling of hot gas in dark matter potential wells. The cooling
in galaxies predominately occurs in the extreme and far UV and so is
not readily observable, but is directly observable in clusters.
Whatever is stemming the cooling in clusters may be determining the
upper mass cutoff for galaxies.

We review these properties of clusters here with an emphasis on our own recent
results.

\section{Cosmological parameters from clusters}

\begin{figure}
\vspace{0.2cm}
\hbox{
\hspace{0.0cm}\psfig{figure=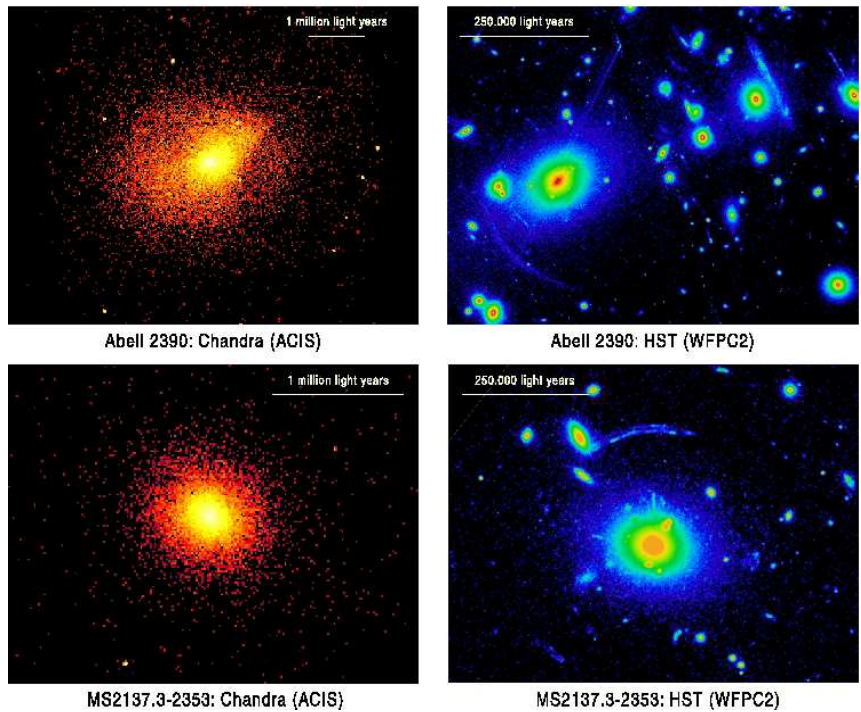,width=0.99
\textwidth,angle=0}
}
\vspace{0.2cm}
\caption{Chandra X-ray (left) and Hubble Space Telescope Wide Field
Planetary  Camera 2 optical (right) images of two of the
dynamically-relaxed, X-ray luminous lensing clusters discussed
here. The clusters shown are Abell 2390 ($z=0.230$) and  MS2137.3-2353
($z=0.313$).  The scale bars indicating distances of 1 million light
years  correspond to angular sizes of 83 and 67 arcsec for Abell 2390
and MS2137.3-2353, respectively. (A standard $\Lambda$CDM cosmology
with $h=H_0/100\kmpspMpc=0.7$, $\Omega_{\rm m}=0.3$ and
$\Omega_\Lambda=0.7$ is assumed.)  Note the clear gravitational arcs
in the HST images.}
\label{fig:montage} 
\end{figure}

The equation of hydrostatic equilibrium $dP/dr=-\rho g$, where $n, T,
P(=nkT), \rho(=n\mu) m$ and $g$ are the particle density, temperature, pressure,
mass density and local gravitational acceleration ($G(M<r)/r^2$), can
be rewritten as
\begin{equation}
M_{\rm T}(<r)=-{{kTr^2}\over{G\mu m}}\left({{d\ln n}\over{dr}} +{{d\ln T}\over
{dr}}\right).
\end{equation}
The quantities on the right can be measured from X-ray spectral images
yielding the gas mass profile $M_{\rm gas}(r)$, the total mass
profile $M_{\rm T}(r)$ and the gas fraction $f_{\rm gas}$. Where
possible, $M_{\rm T}(<r)$ can be checked against gravitational lensing
data (Fig.~1).
$\Omega_{\rm M}$ is then determined using the baryon mass density
$\Omega_{\rm b}$ by [50]
\begin{equation}
f_{\rm gas}={M_{\rm gas}\over M_{\rm T}}={\Omega_{\rm b}\over
\Omega_{\rm M}}.
\end{equation}

\begin{figure}
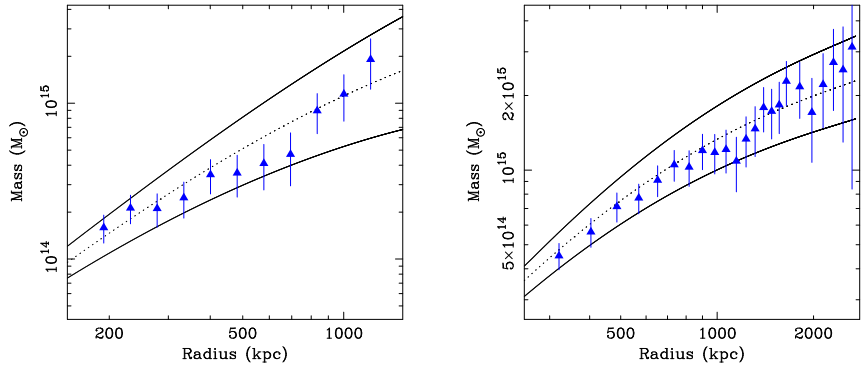

\vspace{0.2cm} \hbox{
\hspace{0.0cm}\psfig{figure=swallen_fig1a.ps,width=0.46\textwidth,angle=270}
\hspace{0.7cm}\psfig{figure=swallen_fig1b.ps,width=0.46
\textwidth,angle=270} } \vspace{0.2cm}
\caption{A comparison of the mass measurements obtained from Chandra
X-ray observations (solid lines) and wide field weak lensing studies
(triangles) of two of the dynamically relaxed clusters in our sample:
Abell 2390 (left; [3]) and RXJ1347.5-1145 (right; [6]).  Error bars
are 68 per cent confidence limits.}
\label{fig:fig1}
\end{figure}

A good example is shown by our analysis of the Chandra data on the
most luminous cluster known RXJ1347-1145 [6]. The X-ray
emission is sharply peaked on the dominant cluster galaxy. Excluding
one quadrant which is hotter due to a merging subcluster, we find that
the X-ray surface brightness and deprojected temperature profiles are
well fit by a hydrostatic model assuming a Navarro-Frenk-White [31]
mass profile (Fig.~2). This also agrees with the redshifts of the strong
lensing arcs and with
weak lensing results. Repeating this for several other massive relaxed
clusters shows that the mass profiles are reliably
obtained from the X-ray data.

We have measured $f_{\rm gas}$ for 9 massive, relaxed clusters out to
the radius where the mean enclosed density is 2500 times the critical
mass density of the Universe at the cluster redshift (Fig.~3; [5]).
This leads to
\begin{equation}
\Omega_{\rm
m}={{(0.0205\pm0.0018)h^{-0.5}}\over{(0.064\pm0.002)(1+0.19h^{0.5})}}
= 0.325\pm0.034,
\end{equation}
where the numerator involves the determination of $\Omega_{\rm b}h^2$
from deuterium abundance measurements in quasar intergalactic
absorption spectra [33] and the Hubble constant
$h=H_0/100=0.72\pm0.08$ [24]. The second small factor
in the denominator accounts for the baryons in stars.

\begin{figure*}
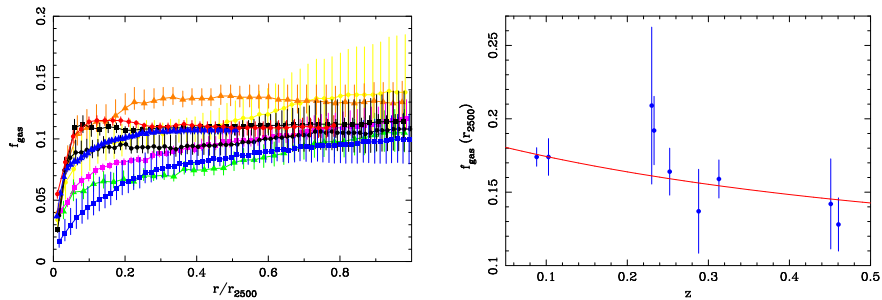

\hbox{
\hspace{0.0cm}\psfig{figure=fgas_r2500.ps,width=0.47
\textwidth,angle=270}
\hspace{0.5cm}\psfig{figure=fgas_z.ps,width=0.49 \textwidth,angle=270}
}
\caption{(Left panel) The X-ray gas mass fraction, $f_{\rm
gas}$, as a function of radius (scaled in units of $r_{2500}$) for the
present sample of nine dynamically relaxed clusters observed with
Chandra [7]. A $\Lambda$CDM cosmology with $h=0.7$ is
assumed. Note how the profiles flatten and converge to universal value
within $r_{2500}$. (Right
panel) The apparent redshift variation of the X-ray gas mass fraction
measured at $r_{2500}$ (with rms $1\sigma$ errors) for a reference
SCDM ($h=0.5$) cosmology. The solid curve shows the predicted $f_{\rm
gas}(z)$ behaviour for the best-fitting, underlying cosmology with
$\Omega_{\rm m}=0.29$ and $\Omega_{\Lambda}=0.68$.  }
\end{figure*}

Numerical simulations of clusters show that $f_{\rm gas}$ should be
independent of redshift. A small amount of gas may be lost during the
assembly of a cluster (remaining fraction $0.93\pm 0.05$; [8]) but
$f_{\rm gas}$ should basically be a universal `standard measure'. We
can therefore use this fact to determine the correct cosmological
model [42, 34], i.e. adjust the cosmological parameters
until we find $f_{\rm gas}$ to be independent of redshift. We find a
best fit with $\Omega_{\rm m}=0.292^{+0.040}_{-0.036}$, $\Omega_{\rm
\Lambda}=0.68^{+0.42}_{-0.52}.$ The method essentially involves the
determination of the angular diameter distance of the cluster $D_{\rm
A}$ ($f_{\rm gas}\propto D_{\rm A}^{3/2}$).

\begin{figure}[ht]
{\includegraphics[width=.5\columnwidth]
{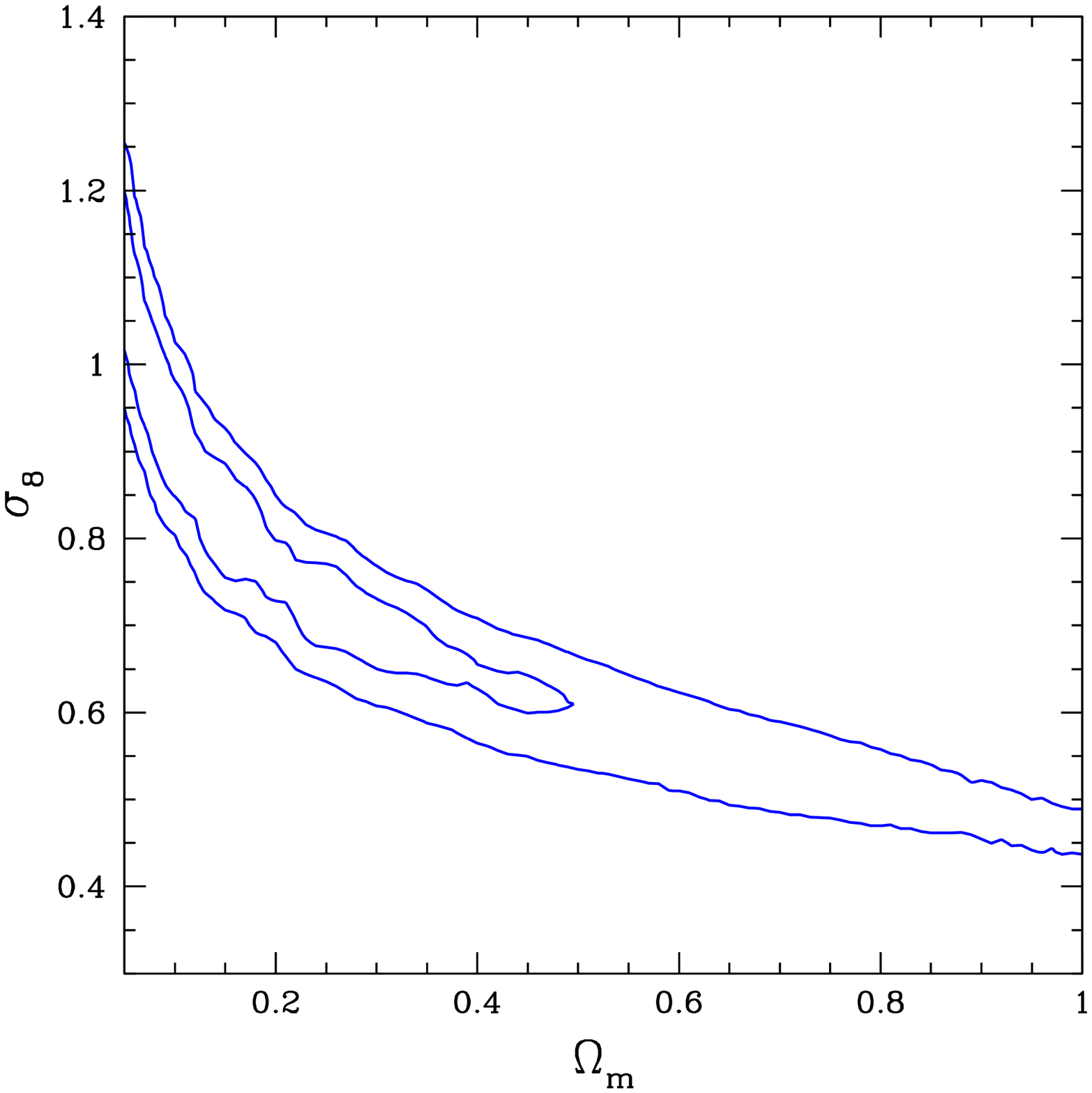}\hfil
\includegraphics[width=.5\columnwidth]
{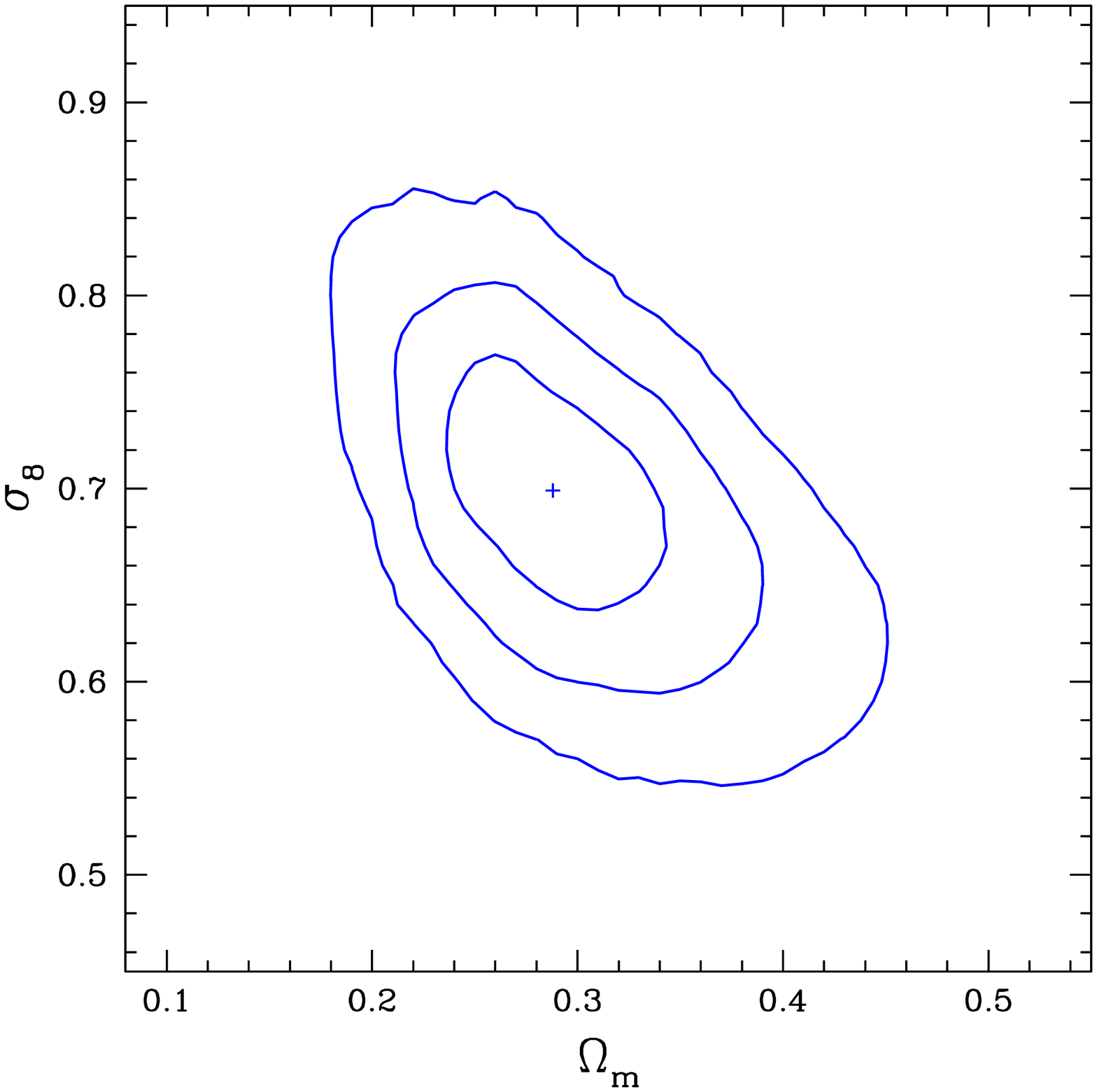}}
\caption
{Left: Allowed region in ($\sigma_8,\Omega_{\rm m}$) space
obtained by applying our mass-temperature relation to the
luminosity function of the 120 ROSAT clusters with luminosities above
$10^{45}\ergps$. Right: The resultant region when our constraints on
$\Omega_{\rm m}$ from $f_{\rm gas}$ are included. }
\end{figure}

The constraints on $\Omega_{\rm m}$ are competitive with all other
methods. Those on $\Omega_{\rm \Lambda}$ show the potential for now.
With another 10 clusters probing the redshift range 0.3--0.8, $f_{\rm
gas}$ measurements promise to exceed the precision for $\Omega_{\rm
\Lambda}$ obtained so far by distant supernovae. The importance of
clusters is that they complement supernovae and sample the redshift
range over which $\Lambda$ is most effective.

We have also determined the mass--luminosity relation from 17 massive
clusters from Chandra observations and weak lensing measurements. In
detail the relation is of $M_{200}$ and $L$(0.1--2.4~keV), the latter
quantity matching the ROSAT band for which there are now excellent
X-ray luminosity functions from the ROSAT All Sky Survey (the eBCS and
REFLEX studies). We then use the combined luminosity function for the
120 clusters with $L$(0.1--2.4)$>10^{45}\ergps$ to obtain the cluster
mass function. This is then compared with the predicted mass function
from the Hubble Volume simulations [18] to yield $\sigma_8$, the rms
variation of the present day, linearly evolved, density field,
smoothed by a top-hat window function of size $8h^{-1}\Mpc$.

The resulting constraint on $\sigma_8$ is a function of $\Omega_{\rm
 m}$ (Fig.~4). If we now include the $f_{\rm gas}(z)$ data the degeneracy
between $\sigma_8$ and $\Omega_{\rm m}$ is broken and we obtain
$\sigma_8=0.695\pm0.042$ [7]. 

Other important and complementary X-ray cluster work on $\sigma_8$ is
by Schuecker et al [44] who have also measured the cosmic power
spectral shape from the distribution of the REFLEX clusters on the
sky, and by Borgani et al [12] who studied the evolution of the
cluster luminosity function.

\section{Cluster cores}

The radiative cooling time within the inner 100~kpc of most cluster
cores is less than $10^{10}\yr$. The gas temperature also drops there
by a factor of two to three (Fig.~5). If there is no heating of the gas it
should cool out at a rate given by (see [19] for a review)
\begin{equation}
\dot M={2\over5}{{L\mu m}\over{kT}}
\end {equation}

\begin{figure}[h]
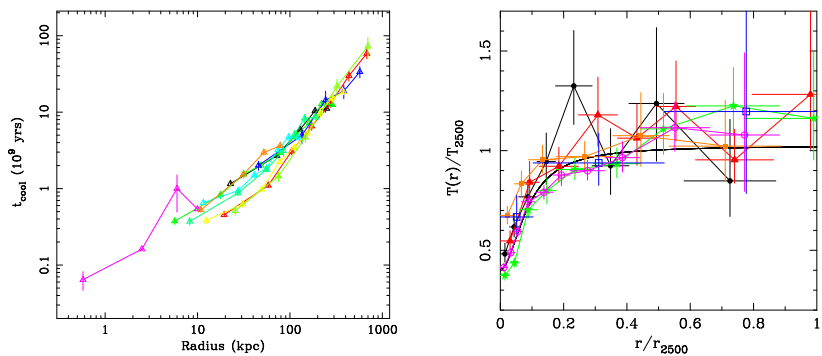

{\includegraphics[width=.4\columnwidth,angle=-90]
{tcool_all2.ps}\hfil
\includegraphics[width=.4\columnwidth,angle=-90]
{temp_profile.ps}}
\caption{Left: Radiative cooling time versus radius for several cooling flow
clusters, from L.~Voigt et al (in preparation). Right: Temperature
profile for 6 massive clusters [4].}
\end{figure}

As the gas cools below 1~keV it emits strong Fe L line emission (e.g.
FeXV emission at 15 and 19~A). A major result from the Reflection
Grating Spectrometer (RGS) on XMM-Newton was to show that little such
emission is seen ([35, 36, 46] Fig.~6). These
studies show that the mass cooling rate below about one half to one
third of the bulk cluster temperature is less than one fifth to one
tenth of that deduced from the above simple formula. Chandra data are
also in agreement with this result. 

\begin{figure}[h]
\includegraphics[width=.6\columnwidth,angle=-90]
{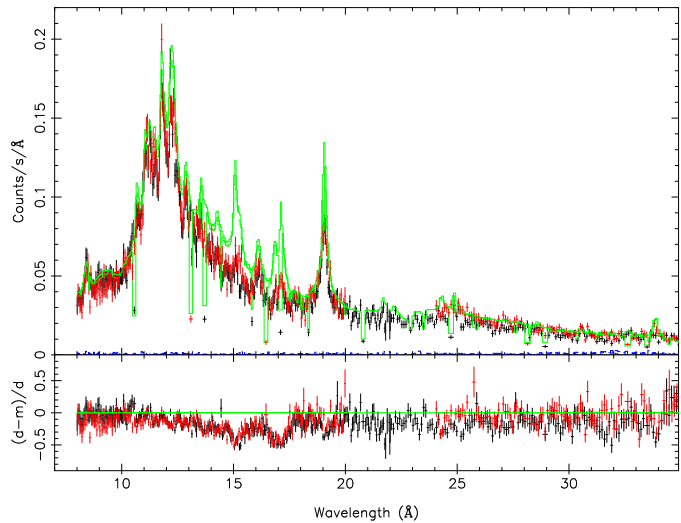}
\caption{RGS X-ray spectrum of the inner parts of the Virgo cluster
around M87 [40]. The faint line shows the predicted
emission if gas is cooling to below $10^6\K$. The data clearly shows
much less emission than predicted by this model in the 13--19~A region. }
\end{figure}

This result was present in previous ASCA and ROSAT studies [2] but the
lack of emission had been attributed to intrinsic absorption. Although
some absorption has not been completely ruled out, the improved new
spectra show that it is not dominant.

It is most likely that some heating is taking place, for which there
are two plausible candidates. These are heating by a central active
nucleus and heating by conduction from the hot outer gas. 

\begin{figure}[h]
{\includegraphics[width=.9\columnwidth]
{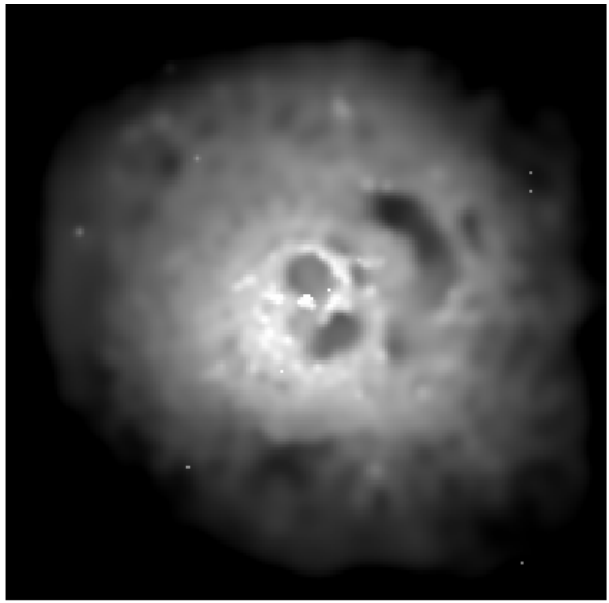}}
\caption{Adaptively smoothed Chandra X-ray image of the core of the
Perseus cluster. Note the holes, which coincide with the radio lobes,
above and below the nucleus. A buoyant outer bubble is seen to the
right. From [20]. }
\end{figure}

All of the relevant {\it cooling flow} clusters peak on a central
galaxy which is expected to host a massive black hole. Many of these
galaxies have radio sources, some of which are obviously blowing
bubbles of relativistic plasma in the central regions (Perseus, [20],
Fig.~7; A2052, [10]).The energy flux from the radio
source can be high ($10^{43-45}\ergps$). The difficulties are however
that the X-ray coldest observed gas lies around the bubbles, not all
clusters host powerful enough radio sources and that the mechanism for
heat transfer from bubbles to the surrounding gas are unclear, despite
several impressive computational studies (e.g. [13, 38, 37]).

The heat must be distributed ([21, 26] Fig.~8) and cannot just heat
the innermost, coolest gas (Fig.~8; [26]). Some [9, 45] have argued that
radio source activity may be sporadic, which explains why there is
little correlation between the present radio source activity and the
heating requirement. However that makes very strong demands on the
power of the source when it is switched on, particularly in the high
luminosity clusters (Fig.~9). There it must exceed $10^{46}\ergps,$
which will hardly be contained in simple bubbles.

\begin{figure}[h]
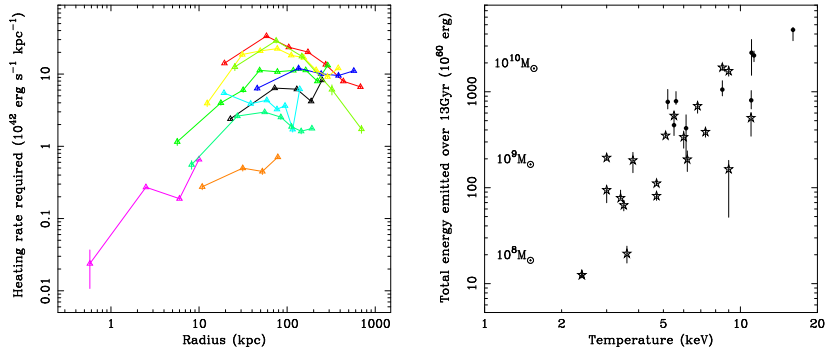

{\includegraphics[width=.4\columnwidth,angle=-90]
{rate_all.ps}\hfil
\includegraphics[width=.4\columnwidth,angle=-90]
{cluster_heat.ps}}
\caption{Left: The heating rate per unit radius required to stop
radiative cooling in several cooling flow clusters (L.~Voigt et al in
preparation). The heating needs to be distributed. Right: The total
energy radiated over a Hubble time from within the cooling region for a
selection of cooling flow clusters. The masses indicate the total mass
which must accrete to produce this energy if the efficiency of energy
release is 0.1. If the central radio source stems cooling in the
hotter clusters then most of the power released must be channeled into
heating the intracluster medium [23]. }
\end{figure}

Note that strong abundance gradients are found in many clusters ([25,
16]), often peaking at radii $\sim 30\kpc$ ([41] Fig.~10; [26]), 
limiting the degree of large scale disturbance which can take place in
the central regions.

\begin{figure}[h]
\vspace{0.2cm} \hbox{
\hspace{0.0cm}\psfig{figure=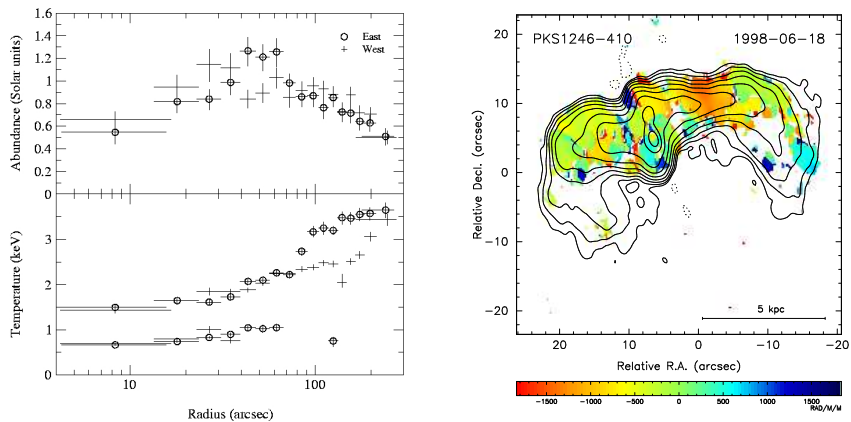,width=0.46\textwidth,angle=0}
\hspace{0.7cm}\psfig{figure=cent_fr.ps,width=0.46\textwidth,rheight=0.55\textwidth,angle=0} } \vspace{0.2cm}

\caption{Left: Abundance (top) and temperature (bottom) profiles for
the Centaurus cluster [41]. Right: Faraday
rotation measure map for the Centaurus cluster [47].}
\end{figure}

Thermal conduction has long been considered to be suppressed in
clusters because of the observed central temperature drops. Conductive
energy flow increases strongly with temperature, unlike radiative
cooling which decreases (at constant pressure), so one might assume
that it either operates, so making the core isothermal, or is
suppressed and radiative cooling dominates. Narayan \& Medvedev [30]
have however revived the concept and noted that conduction may account
for the observed temperature gradients. This has been explored in more
detail by [48, 23, Fig.~11] and [51]. These last authors found some
clusters where conduction appears to be insufficient. A major issue
here is whether the effective conductivity can be as high as the
Spitzer value, or whether magnetic fields suppress it heavily (Narayan
\& Medvedev [30] argue that it may
operate at close to the Spitzer rate). Faraday rotation indicates the
presence of magnetic fields in the intracluster medium ([47] Fig.~10).

Ruszkowski \& Begelman [39] have incorporated both distributed heating
by a radio source with conduction to obtain stable solutions. This
still leaves the method by which the heat is distributed as unsolved.

It is unlikely that either radio source heating or conduction can
suppress radiative cooling within such a large region completely.
Indeed, this is unnecessary, for significant rates of massive star
formation [1, 15] and masses of cold gas [17] are found in the central
parts. It is probable that
\begin{equation}
\dot M_{\rm cool}=\dot M_{\rm X}/10,
\end{equation}
where $\dot M_{\rm X}$ is the cooling rate derived simply from the X-ray data
on the assumption of no heating.

In summary there are plausible heat sources at the centre and beyond
radii of 100~kpc. The main problem is to distribute the energy within
100~kpc without either disrupting the metallicity profiles or
exceeding some observational constraint. Beyond any bubbles and
plumes, and an occasional cold front (all of which occupy only a small
fraction of the volume of the cooler gas) the distribution of surface
brightness, temperature, metallicity and entropy of the gas all vary
very smoothly. 

Further possibilities remain in which cooling dominates but the
situation is more complicated so that gas cooling below say 2~keV less
observable. This can result if the metals are not uniformly mixed in
the hot gas [21, 29] or if the cooler gas mixes with cold gas [22].
The missing soft X-ray luminosity from a simple cooling flow is
similar to that in the optical/UV/IR nebulosity at the centre. One
reason to continue to consider such models is the detection of strong
OVI emission in some clusters [32].

\begin{figure}[ht]
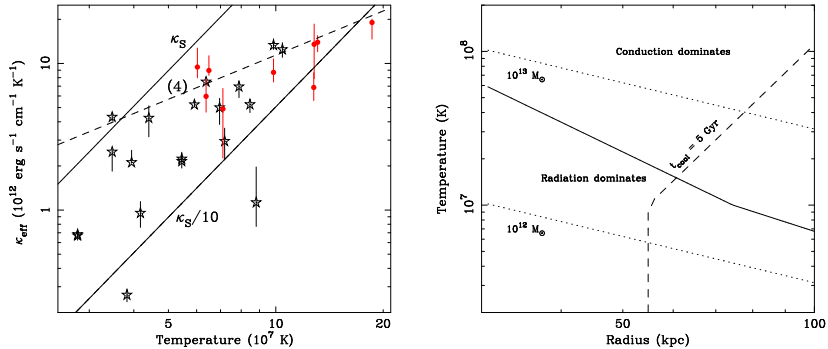

{\includegraphics[width=.4\columnwidth,angle=-90]
{kappaeffplot_acf.ps}\hfil
\includegraphics[width=.4\columnwidth,angle=-90]
{galform.ps}}
\caption{Left: Effective conductivity required to balance radiative
cooling in a sample of the brightest clusters [23].
Right: Conduction can only offset radiative cooling in very massive
galaxies, and so may determine the upper mass limit of galaxies [23]. }
\end{figure}

The cooling flow problem, as it has become known, has wider relevance
than just to cluster cores. The visible parts of galaxies are due to
gas cooling in dark matter potential wells [49, 28] and the cores of
clusters are a directly observable example of this process. If it does
not operate in cluster cores why does it work in galaxies? It is
possible that whatever is stemming cooling in clusters does so in
galaxies but operates in such a way as to dominate only in massive
systems, so determining the upper mass limit of visible galaxies. A
process like conduction, which is more effective in hotter, massive
objects, has the right property to allow gas to cool in normal
galaxies but not in more massive systems ([23] Fig.~11).

Although cluster cores are complex regions the superb spatial
resolution of Chandra means that it can be separated from the bulk of
the cluster and need not affect the cosmological determinations
outlined earlier.

\section{Acknowledgements}

We thank Robert Schmidt and Jeremy Sanders for help and collaboration
with these projects.

\end{document}